# A comparative study on defect estimation using XPS and Raman spectroscopy in few layer nanographitic structures

K. Ganesan,[*a] Subrata Ghosh,[a,b] Nanda Gopala Krishna,[c] S. Ilango,[a] M. Kamruddin[a,b] and A.K.Tyagi[a,b]

[a] Materials Science Group, Indira Gandhi Centre for Atomic Research, Kalpakkam - 603102, India.

[b] Homi Bhabha National Institute, Anushaktinagar, Mumbai - 400 094, India

[c] Corrosion Science and Technology Group, Indira Gandhi Centre for Atomic Research, Kalpakkam - 603102, India.

* Corresponding author : kganesan@igcar.gov.in

## Abstract

Defects in planar and vertically oriented nanographitic structures (NGSs) synthesized by plasma enhanced chemical vapor deposition (PECVD) has been investigated using Raman and X-ray photoelectron spectroscopy. While Raman spectra reveal the dominance of vacancy and boundary type defects respectively in vertical and planar NGSs, XPS provides additional information on vacancy related defect peaks at C 1s spectrum that originate from non-conjugated carbon atoms in hexagonal lattice. Although an excellent correlation prevails between these two techniques, our results show that estimation of surface defects by XPS is more accurate than Raman analysis. Nuances of these techniques are discussed in the context of assessing defects in nanographitic structures.

**Introduction**

In the recent past, graphene and related materials have attracted much attention in the arena of electronic and optoelectronic devices, chemical sensors, biosensors, and electrochemical energy storage devices due to their exotic structural, electronic and optical properties.[1,2] Apart from monolayer graphene, its variant few layer graphene (FLG) is also found to exhibit similar interesting properties with more structural stability.[3,4,5] Further, it has immense resistance even to harsh chemical environments and heavy ion irradiation. Thus, devices based on FLG are expected to be more stable than monolayer graphene, wherein the defects influence their performance.

Defects prevail invariably in all solids and generally considered to be detrimental to the devices based on such materials. Nevertheless, the defects in graphene are rather interesting and can be used to control its electronic, optical, magnetic and mechanical properties. For example, the conductivity of graphene increases with increasing vacancy defects by more than one order of magnitude in contrary to the normal expectation.[6] Apart from this, band gap can also be introduced in graphene by choosing appropriate doping impurities or by varying the width of graphene nanoribbons.[7] Further, room temperature ferromagnetism in graphene is exhibited by strain, hydrogenation, vacancies, and edge defects.[8,9] Thus, the defects and disorder found to play a major role in determining the physical and mechanical properties of graphene-like materials. Hence, it is essential and important to have a thorough knowledge on characterization of defects that are created in nanographitic structures (NGSs) during growth process.

In this regard, an array of analytical tools are available to characterize the defects in carbon materials viz. Raman spectroscopy (RS), Fourier transform infrared spectroscopy, XPS, positron annihilation spectroscopy and scanning tunneling microscopy. Among them, RS is the most versatile technique owing to its simplicity, non invasive nature and the potential to analyze the defects and disorder in carbon materials.

Raman spectroscopy is extensively used for investigating structural disorder in carbonaceous materials based on various types of defects namely point defects, edges, doping, stacking fault, grain boundaries, stress, and strain.[10,11,12,13] In addition, the degree of disorder in carbonaceous materials are also measured by qualitatively and quantitatively using RS.[14,15,16,17]- In spite of these advantages, RS has certain inherent limitations in detecting defects in these materials like charged impurities, perfect zigzag edges, intercalants, uniaxial and biaxial strain, as they do not manifest as defect band (D band).[12] Furthermore, the defects caused by surface-bound functional groups also do not leave any significant footprints in Raman spectrum.[18] These aspects of RS introduce difficulties in estimating the defect density, particularly in FLG and hence motivate the current work to investigate these problems using XPS as an alternative tool.

XPS is a well known and powerful technique to study the surface chemical properties of materials. In addition to ascertain chemical composition and bonding environment, it can also provide useful information about defects and disorder in carbonaceous materials. The presence of defects and functional groups in these materials introduce new peaks and alter the peak position and line width of C 1s band. In particular, it can quantify the amounts of C-C bonding ( $sp^1$, $sp^2$ or $sp^3$ ), doping, and reveal other functional groups. Recent experiments and theoretical calculations on thermally/chemically oxidized carbonaceous materials have proved the ability of XPS technique in accurate estimation of defects in carbonaceous materials.[19,20,21,22,23,24] In this respect many reports do exist on analysis using XPS, but most of them are devoted to studies on oxidized graphite, carbon nanotubes, and fullerenes.[25,26,27,28,29] However, it is found only very few of them have reported on measurement of defect density in FLG using XPS. These earlier studies using XPS have indeed recorded a characteristic features pertaining to defects in graphitic structures with a new broad band (~ 284 eV) at slightly lower binding energy (BE) of C-C $sp^2$ bonding. But this band is completely ignored initially by several researchers and later it is proved that this band originates from vacany-like defects in graphitic lattice. In the present study, we effectively utilize this vacancy related defect band in addition to the non-conjugated carbon band which arises at B.E. of 285eV to compare with results obtained from RS. This study further investigate to extend the scope of defects estimation by a combined use of RS and XPS which is rarely ever reported in literature. Hence, this study is devoted to extend and overcome the inherent limitations posed by RS in defect analysis using XPS as an alternative tool in the determination of defect density in FLG grown by PECVD.

Among the variety of techniques to synthesis FLG, the plasma enhanced chemical vapor deposition (PECVD) holds the promise to grow NGSs in a controlled manner with large area, conformal and uniform deposition having numerous advantages. Also, it is important to note that the synthesis is catalyst free and can be grown on any type of substrates such as metals, semiconductors and insulators.[30] Thus, PECVD avoid a major hurdle of transferring the graphene onto a dielectric substrate for device fabrication. Further, this technique enables to grow NGSs consists of a few layers of graphene with varying morphology and orientation.[31] However, with all these advantages to its credit, NGSs synthesized by PECVD are also prone to large number of defects caused by hydrogenation, large grain boundaries, edges, atmosphere adsorbents and ion bombardment due to the fast growth rate that takes place in high density plasma under non-equilibrium conditions. Thus, the study of defects also aids in optimizing the deposition parameters.

For the present study four different samples are prepared based on our previous study with different feed stock composition.[31] These varying gas compositions are proved to yield NGSs with defects of varying proportions. The very fact that, these samples do not have any intentionally added functional groups other than adsorbed molecules in very small concentration from atmosphere. While cursory characterization of these samples are carried out using SEM, AFM, and Hall measurements, a detailed investigation is made using XPS and Raman. In addition, a comparative study on defects estimated by these techniques is made to elucidate how XPS can effectively address the finer aspects of quantification of defects in FLG.

**Experimental methods**

NGSs on SiO2/Si substrates were grown under different feedstock gas compositions using PECVD. The details on experimental set up and deposition chamber were published elsewhere.[31] For the present study, samples were prepared with varying defect density through four different set of trials by choosing appropriate feedstock ratios of $CH_4$:Ar:$H_2$ as 1:0:0, 1:0:5, 5:3:1 and 1:5:0 and the films are labeled as S1, S2, S3 and S4 respectively. All the samples were grown on $SiO_2$/Si substrates heated to $800^0C$ using a vacuum compatible heater. For deposition 400W of microwave power was used and the deposition time span around 30 minutes. After deposition, plasma was turned off and the samples were further allowed to remain at growth temperature for another 30 minutes. Samples are then brought to room temperature by switching OFF the heater.

Subsequently, the samples were characterized for their morphological and structural features using field emission scanning electron microscope (SEM) and atomic force microscope (AFM). The degree of graphitization in terms of defects and disorder were evaluated by micro-Raman spectroscopy (inVia Renishaw, UK) in the back scattering geometry. The spectrometer is equipped with $Ar^+$ ion laser (514 nm) as the probe source with grating monochromator (1800 grooves/mm) and Peltier cooled charge coupled device as detector. A microscope objective of 100X magnification with numerical aperture of 0.8 was used in this study. To avoid the laser induced heating on samples, laser power was kept below 1 mW. The acquired spectra were analyzed using WIRE3.2 software and fitted with Lorentzian line profiles. Estimation of defects are obtained using peak intensity ratio of Raman bands D and G ($I_D/I_G$).[32]

Surface chemical properties in NGSs were characterized using XPS (M/s SPECS, Germany) system equipped with a monochromatic Al Kα source (1486.7 eV) operated at 350W with detection pass energy of 10 eV. Refined spectra for all the samples after Shirley background subtraction were fit with Gaussian–Lorentzian line shapes with fairly similar fitting parameters. Hall measurements were made in van der Paw geometry with Agilent B2902A precision source/measure unit to estimate sheet resistance, mobility and carrier concentration.

**Results and discussion**

**Surface morphology**

Fig. 1 (a-d) shows the cursory view of the morphological features NGSs obtained through SEM for samples grown under different feedstock ratios. A closer view of these features is examined using an AFM and their corresponding topography images are given in Fig. 1 (e to h). As can be seen from these images, the surface morphological features of the grown NGSs (S1-S4) vary significantly under different feedstock ratios. While AFM images show the finer aspects of topographical features of the samples S1, S2 and S3, SEM provides a better representation on the 3D structure of sample S4. Among them, S1 and S2 have nearly similar features and exhibit high density of isolated spherical carbon nanoclusters. Herein, in S2 the aerial density of isolated spherical carbon nanoclusters is less when compared to S1 as evidenced from Fig.1f. The sample S3 shows very smooth surface morphology as shown in Fig 1c and Fig 1g. On the other hand, S4 shows dense *3D* networks of vertically oriented graphene nanosheets with petal-like structures as depicted in Fig.1d. The corresponding AFM image of this sample is found to be different from SEM micrograph. This is possible due to the convoluted effect of large radius of curvature and short tip height of the cantilever used in AFM. Thickness of these samples grown under different feedstock compositions is found to be 12.1, 4.3, 17.5 and 136.8 nm for the samples S1, S2, S3 and S4 respectively, as estimated from cross sectional SEM and AFM. The root mean square (rms) roughness is found to be 3.0 and 1.4 nm for the samples S1, and S2 respectively while the sample S3 consists of extremely smooth surface with rms roughness of 0.5 nm, measured over an area of 1 x 1 $\mu m^2$. The rms roughness of S4, vertical NGSs is found to be 39.3 nm.

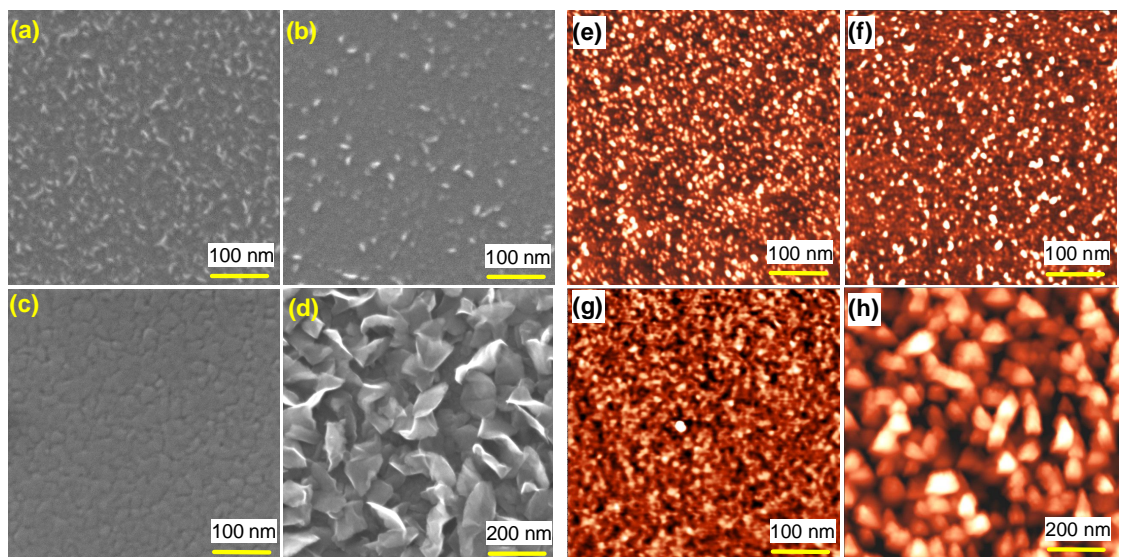

Fig. 1. SEM micrographs and their respective AFM topography features of nanographitic structures grown under different feedstock compositions of $CH_4:Ar:H_2$ ratios as (a,e) 1:0:0, (b,f) 1:0:5, (c,g) 5:3:1 and (d,h) 1:5:0.

SEM and AFM analysis indicate that morphology of these samples varies from perfect planar (S3) to homogeneous 3D networks of vertical graphene sheets (S4) with intermediate isolated carbon nanoclusters (S1 and S2) that depends on feedstock composition. In general, S1, S2 and S3 can be considered as planar NGSs with large point defects and grain boundaries while S4 can be considered as a bunch of vertical graphene sheets with large number of edges.

The growth mechanism of NGSs is fairly understood in PECVD synthesis, in which the growth orientation is always initiated from the nanocrystallites of carbon at the base layer. This layer is a sacrificial layer which evolves at the very early stages of nucleation over a heterogeneous substrate surface just before the onset of vertical or planar orientation. Such a base layer consists of high aggregation of carbon with defects structure having $sp^2$ and $sp^3$ bonding due to their fast and non-equilibrium growth during deposition processes. The thickness of base layer generally depends on the growth condition and can vary up to a few tens of nanometers. The competing process of growth and etching greatly dictates the morphology of the NGSs by introducing varying defect concentration. Generally, it is found that Ar-rich feedstock composition favors vertically oriented graphene nanosheets with lesser defects whilst $H_2$-rich composition aids growth of planar NGSs with higher defect concentration. The combined dilution of $CH_4$ with Ar and $H_2$ give rise to a change in the growth orientation and in turn provides a handle to tune the defect concentration. [31]

**Raman spectroscopy**

Fig.2 shows the Raman spectra for the samples under investigation. These spectra consist several Raman bands such as D, G, D′, G′ (also called as 2D) and D+D′ bands which are typical for a defective nanographitic system.[33] The presence of large D band intensity and intra-valley defect band (D′) in theses samples confirm the existence of defects and disorder. Herein, samples S3 and S4 have a distinct shoulder of D′ in G band that indicates existence of disorder at a lower scale while the samples S1 and S2 have overlapped graphitic G and D′ bands representing significant disorder. As can be seen in Fig.2a, high D band intensity is observed in theses samples and it is generally attributed to the dominant $sp^3$-hybridized C–C bonds which may arise from grain boundaries, vacancies, pentagons, heptagons and graphene edges.[33]

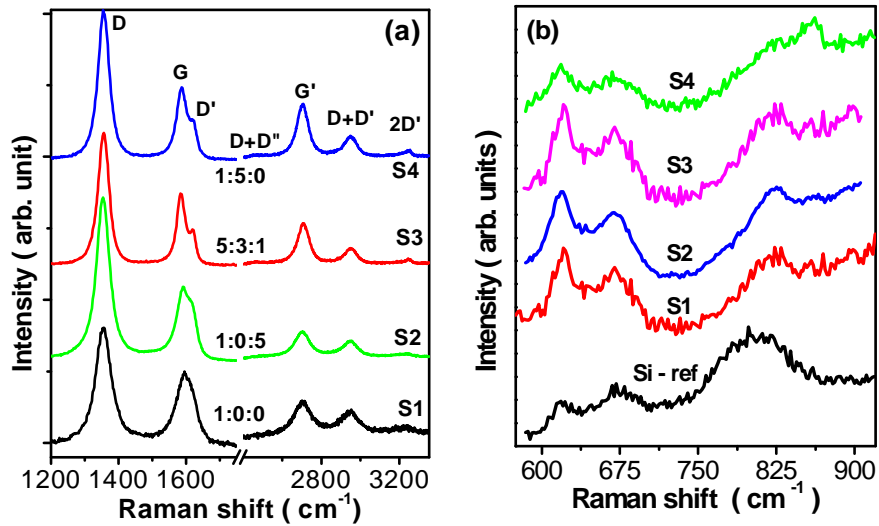

Fig.2. (a) Raman spectra of nanographitic structures grown under different feedstock compositions of $CH_4$:Ar:$H_2$ ratios as 1:0:0, 1:5:0, 5:3:1 and 1:0:5 (b) A closer view of Raman spectra with silicon reference.

Table 1: The extracted parameters from Raman and Hall analysis of nanographitic structures grown at different feedstock compositions

| Sample | Feed-stock | height (nm) | Peak position (cm$^{-1}$) | | | FWHM (cm$^{-1}$) | | | $I_D/I_G$ | $I_D/I_{D'}$ | $L_a$ (nm) | $\rho_{sh}$ (k$\Omega$/sq.) | p (10$^{13}$ cm$^{-2}$) | $\mu$ (cm$^2$/Vs) |
|---|---|---|---|---|---|---|---|---|---|---|---|---|---|---|
| | | | D | G | G' | D | G | G' | | | | | | |
| S1 | 1:0:0 | 12.1 | 1355.4 | 1592.4 | 2703.5 | 57.9 | 57.3 | 109.0 | 1.79 | 5.0±0.6 | 1.8±0.07 | 4.2 | 36.8 | 4.0 |
| S2 | 1:0:5 | 4.3 | 1354.2 | 1590.2 | 2702.9 | 46.8 | 45.2 | 88.4 | 2.49 | 4.4±0.5 | 6.8±0.07 | 8.6 | 6.1 | 12.0 |
| S3 | 5:3:1 | 17.5 | 1355.3 | 1584.8 | 2707.4 | 38.2 | 29.3 | 69.7 | 1.75 | 4.8±0.3 | 9.6±0.26 | 4.4 | 2.3 | 63.8 |
| S4 | 1:5:0 | 136.8 | 1355.9 | 1586.7 | 2704.1 | 39.7 | 36.7 | 69.9 | 2.09 | 5.4±0.4 | 8.0±0.03 | 1.0 | 112.6 | 5.6 |

The intensity ratio of Raman bands D and G ($I_D/I_G$) can reveal the amount defects in carbon materials.[12] Table 1 lists the extracted parameters of the best fit from Raman spectra using Lorentzian line shape analysis. Although samples S1 and S3 have nearly same $I_D/I_G$ ratio, S1 has a larger FWHM than S3. Hence, quantification is made by taking into account of FWHM in combination with $I_D/I_G$ ratios. Raman analysis on samples reveals that the defects are governed by a three stage model, known as "amorphization trajectory of graphite" proposed by Ferrari and Robertson.[34] According to this model, $I_D$ is directly proportional to the defect concentration in stage 1 and inversely proportional in stage 2. In addition, $I_G$ is directly proportional to the number of sp$^2$ rings and chains in the sample.[35, 36] As per this model, sample S1 is governed by stage 2, whereas all other samples follow stage 1. In all these NG structures, the origin of D band arises mainly from the surface edges and exhibits a one dimensional character akin of zig-zag and arm-chair edges in graphene. Thus, the extent of disorder in NG can be quantified by the span of boundaries with respect to total crystallite area. This aspect can also be defined in terms of inverse nanocrystallite size ($L_a$).[36] The $L_a$ in NGSs can be appropriately calculated using T-K relations as given below. [16, 37]

$$L_a(nm) = \frac{560}{E_L^4}\left(\frac{I_D}{I_G}\right)^{-1} \qquad (1)$$

$$L_a(nm) = \left[\frac{1}{c(\lambda)}\left(\frac{I_D}{I_G}\right)\right]^{\frac{1}{2}} \qquad (2)$$

where, $E_L$ is the energy of the laser in eV and $C(\lambda)$ is a constant which is 0.55 nm$^{-2}$ for 514.5 nm laser. Equations 1 and 2 respectively valid for low (stage 1) and high (stage 2) defect concentration regimes in the graphitic structures. The calculated $L_a$ for these samples, S1, S2, S3 and S4, are found to be 1.8, 6.8, 9.6 and 8.0 nm respectively. Large $L_a$ value implies better crystallinity in the system. Thus, the samples from S1 to S4 tend to have a monotonic decrease in defect density. In other words, the crystallinity of NGSs improves upon dilution of $CH_4$ with hydrogen (S2) and then further improves under argon (S4) and for Ar+H2 gas mixture (S3) as evidenced from the increase in $L_a$.

As can be seen in Table 1, the position of D band for all the samples is found at around ~ 1355 cm$^{-1}$. The blue shift in G and 2D bands also confirms the extend of disorder, with respect to pristine graphene peaks at 1582 and 2700 cm$^{-1}$ respectively.[10] As given in table 1, the samples S3 and S1 exhibit blue shift in G band to the extend of 2.8 and 10.4 cm$^{-1}$ respectively and this is in direct correlation with amount of disorder as measured by $L_a$. Blue shift in G band also implies unintentional doping which could arise from defects and adsorbed atmospheric molecules.[10] Further, the nature of doping is found to be due to holes as inferred from the blue shift in the 2D bands.

Further, based on an empirical model, intensity ratio of $I_D/I_{D'}$ can also reveal the nature of defects in graphene.[35] According to the model, the $I_D/I_{D'}$ of 3.5 corresponds to boundary-like defects, 7 represents vacancy-like defects, 10.5 attributes to hopping defects and 13 corresponds to the presence of sp$^3$ related defects. The physical origin of the defects in carbonaceous materials can be accounted as follows : the boundary like defects mainly arise from the scattering of phonons with defects from grain boundaries; the vacancy-like defects represents the single or double vacancy in the graphitic lattice; the hopping defects are any defects that lead to deformations of the carbon-carbon bonds in graphene and sp$^3$ related defects represent the covalent sp$^3$ bonded functional groups on the carbon atoms.[38] In the present study, the $I_D/I_{D'}$ ratios of these samples are found to be about 5. Based on this, the intermediate value of $I_D/I_{D'}$ ratio can be reasonably attributed to the mixed vacancies and boundary type defects. Among the samples, S2 and S4 have the lowest and highest $I_D/I_{D'}$ ratios of 4.4 and 5.4 respectively. Thus, we can conclude from Raman analysis that the sample S4 with vertical graphene orientation is mostly dominated by vacancy- like defects in addition to small boundary-like defects. Similarly, the other planar nanographite samples S1, S2 and S3 have dominant boundary-like defects along with small vacancy-like defects.

Hall measurements performed on these samples indeed reveal that the systems behave as p-type material. Their sheet resistance ($\rho_{sh}$), carrier concentration (p) and mobility (µ) are listed in Table 1. Apart from this, adsorbed –OH functional group on basal plane / edges of NGSs can also manifest as p-type doping. Since sample S4 has large 3D network of graphene edges with atmosphere adsorbent, it contains highest hole concentration and low mobility. But sample S3 exhibits the highest mobility among all samples and indicate the best structural quality. This observation based on Hall measurements is consistent with defect analysis by RS.

**X-ray photoelectron spectroscopy**

Fig.3 shows the C 1s spectra for the four samples (S1, S2, S3 and S4). The spectra show the graphitic nature of the samples with their typical asymmetry in the higher binding energy (B.E) regime. The best fit for the main C1s spectrum is obtained by deconvoluting the profile into six Gaussian–Lorentzian line shapes and their respective parameters are presented in Fig.3a-d. Table 2 provides the best fit parameters used for the XPS analysis. The prominent peak at 284.4 eV is assigned to $sp^2$ C=C bonds and other low intensity peaks at 285.6, 286.3 and 287.0 eV are attributed to alcohol (C-OH), ether/epoxy (C-O-C) and ketone/aldehyde (C=O) functional groups on the surface respectively.[19] Herein, it is to note that graphitic structures heat treated to high

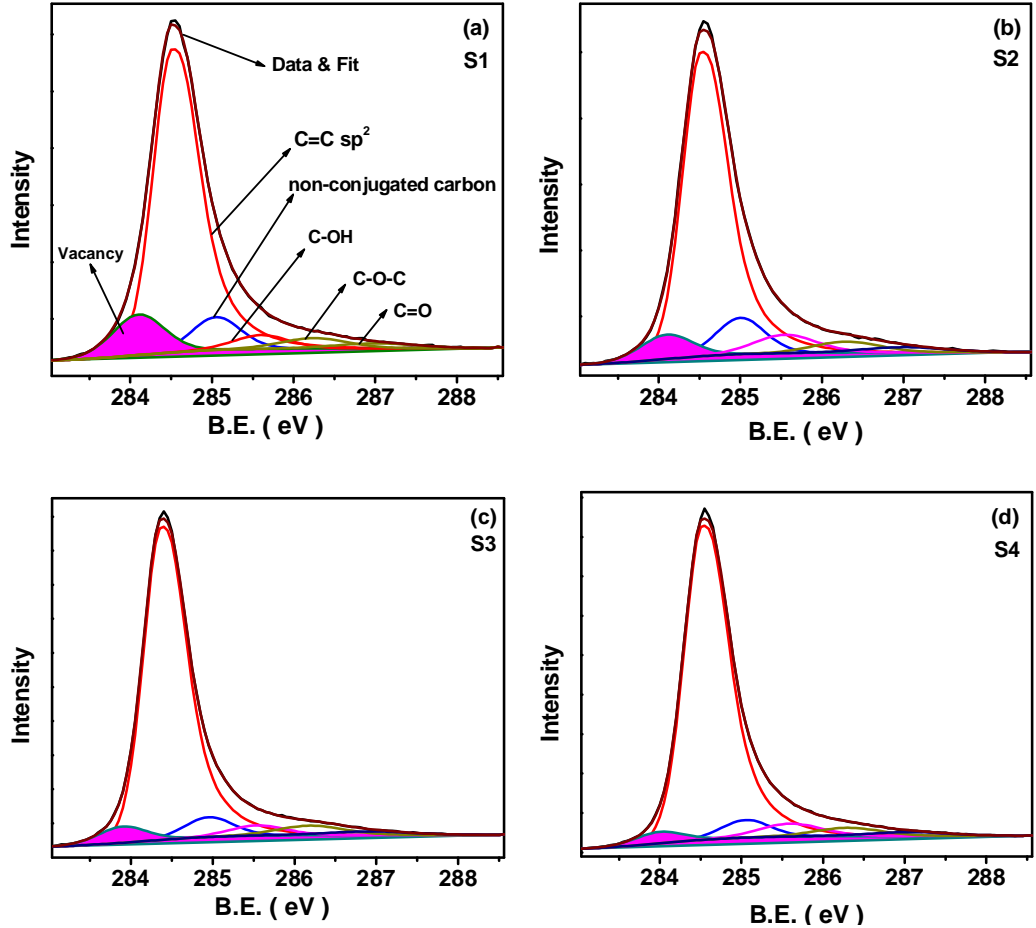

Fig.3. XPS spectra of C1s and its deconvoluted profiles for films grown under feedstock compositions of $CH_4:Ar:H_2$ ratios as (a) 1:0:0 (b) 1:0:5 (c) 5:3:1 and (d) 1:5:0

temperature are more prone to atmospheric moisture and get adsorbed in ambient conditions. These adsorbed molecules by the highly reactive graphene defect sites such as edges, grain boundaries and point defects manifest themselves as functional groups in XPS spectra. The other two peaks at 284 and 285 eV have some interesting features related to defects in carbon materials and are discussed in detail in the next paragraph. In order to rule out any other origin of the peak at 284.0 eV, which is closely matching to the BE for partial Si-C alloy, very careful observations were conducted to search for Si-C alloy formation at $SiO_2$-nanographite interface using Raman spectroscopy. However, the signature of SiC, which should have been appeared on Raman band at about 800cm$^{-1}$, is found absent as verified from fig 2b. At this juncture, a brief glimpse on the earlier observations reported in literature related to defects in carbonaceous materials are noteworthy in the context of our results and discussion.

Historically, the occurrence of broadening in C1s line shape of carboneous materials was first reported by Szwarckopf. [23] According to this report the presence of defects shows up as a peak at 285.2 eV in the XPS spectrum. Further it was accounted that the peak broadening is in direct proportion to the defect concentration. Since then, several works on defective carbon materials have mentioned about this peak. Later, Barinov et al [19] had proved through experiments and DFT calculations that the line broadening and additional peaks arise due to the presence of point defects in hexagonal carbon lattice. According to their calculation, a single C vacancy in hexagonal lattice causes chemical shift of 12 near-by atoms ranging from first to third in-plane nearest neighbors which results in the evolution of defect bands at lower BE regime of C 1s band. Such broadening is further accounted with four additional peaks stemming from the presence of carbon vacancy in different environment at the lattice site. Further, the C 1s line shape broadens dramatically when the defect density becomes very large. To fit the large line broadening, an additional peak corresponding to BE of 285 eV become necessary and this peak is attributed to the presence of C adatoms or sp3-C bond over the surface. These vacancy related defect peaks are also experimentally verified by creating vacancy-like defects using Ar ion bombardment on highly oriented pyrolytic graphite. [19] Also, the peak at 285 eV is assigned to non-conjugated carbon (nc-C) bonding in multi-walled CNTs and C-H bonding in hydrogenated graphite which are all generally termed as defects. [21, 39] In addition, a detailed calculation on the effect of different types of functional groups attached at the edges and basal plane with their corresponding peak positions and FWHM obtained from XPS spectra are carried out by Yamada et al [20] and verified experimentally.

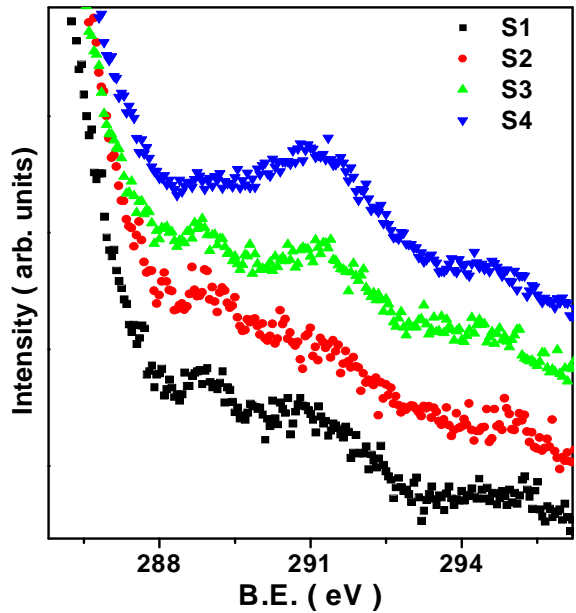

Fig.4. Zoomed part of C1s spectra corresponding to the BE regime of π-π* transition for films grown under different feedstock compositions of $CH_4$:Ar:$H_2$. The spectra are given vertical offset for the shake of clarity in viewing.

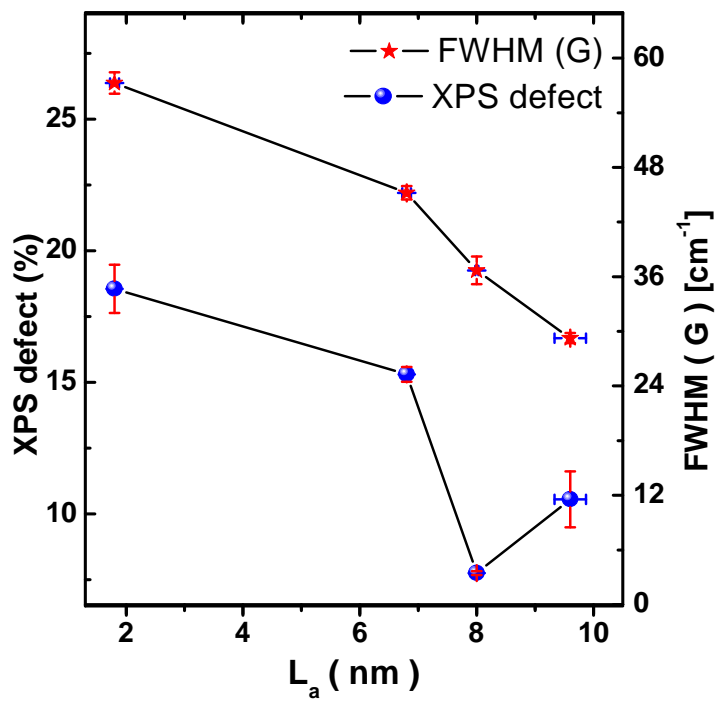

Fig 5. Variation in defect concentration and FWHM of G band obtained from XPS and Raman spectroscopy respectively are plotted as a function of average crystallite size (La) observed in the nanographitic structures. The solid lines in the plot are guide to eye.

Table 2. The parameters extracted from XPS analysis for the nanographite films grown at different feedstock compositions

| Sample | C=C sp$^2$ | | nc-C | | C-OH | | C-O-C | | C=O | | vacancy | | Total defect [%] |
|---|---|---|---|---|---|---|---|---|---|---|---|---|---|
| | B.E. [eV] | Area [%] | B.E. [eV] | Area [%] | B.E. [eV] | Area [%] | B.E. [eV] | Area [%] | B.E. [eV] | Area [%] | B.E. [eV] | Area [%] | |
| S1 | 284.5 | 71.6±2.1 | 285.0 | 10.2±3.5 | 285.6 | 4.6±1.3 | 286.3 | 3.6±0.5 | 287.0 | 1.5±0.6 | 284.1 | 8.3±2.6 | 18.5±0.9 |
| S2 | 284.5 | 73.3±0.7 | 285.0 | 9.1±0.4 | 285.6 | 5.4±0.2 | 286.3 | 4.0±0.6 | 287.0 | 1.9±0.2 | 284.1 | 6.3±0.1 | 15.3±0.3 |
| S3 | 284.4 | 81.7±0.1 | 285.0 | 4.8±0.9 | 285.6 | 3.2±0.1 | 286.2 | 3.1±0.7 | 286.9 | 1.4±0.4 | 283.9 | 5.8±2.0 | 10.6±1.1 |
| S4 | 284.5 | 82.6±0.6 | 285.0 | 4.8±0.1 | 285.6 | 4.4±0.1 | 286.3 | 3.6±0.9 | 287.1 | 1.6±0.3 | 284.0 | 3.0±0.1 | 7.8±0.1 |

In the present study, all the samples exhibit peaks at 284 and 285 eV indicating the presence of defects. As discussed earlier, the peak at around 284.1 eV is attributed to the presence of point defects which could arise from C vacancies, pentagon and heptagon rings, and formation of fullenerene like structures at the early stages of nucleation process and also with functional groups attached to graphitic lattice.[19,20] Further, we attribute the peak at 285 eV to the nc-C in the hexagonal lattice which is due to combination of C adatoms, non-aromatic C atoms and hydrogenation of carbon on the surface of NGSs. As shown in table 2, the sample S1 has the highest total defect concentration of ~ 18% of which 8.3% is from vacancy defects and ~ 10.2% from nc-C defects. Other planar NGSs, S2 and S3, have lower defect concentration with vacancy defects of ~ 6.3 and 5.8 % and nc-C defects of 9.1 and 6.3% respectively. On the other hand, the vertical NGSs (S4) has the lowest defect concentration with vacancy and non-conjugated C defects of about 3.0 and 4.8 % respectively.

Another indication of a structurally superior graphitic material can be inferred from broadened shake-up (ShU) band at the high energy tail of C1s spectrum. The ShU high energy tail takes place through a mechanism, wherein the $\pi$-electrons makes a rapid response to the photo-excitation of core 1s electrons.[21,40] The relative magnitude of this ShU signal can be correlated with the conjugation strength which depends on the amount of C=C sp$^2$ bonding in the hexagonal lattice.[41] Further, it is possible to evaluate the structural quality of the graphitic system using ShU peak. Fig 5 shows the appearance of ShU peak for the samples under investigation in the BE range of 288 to 295 eV. As seen in Fig.5, the relative area under the curve ($A_{ShU}$) increases monotonically from S1 to S4. This reaffirms S4 has a superior graphitic structure with less defects when compared to other samples. In nutshell, the sample S1, grown without any dilution, has the highest vacancy and nc-C defects while S4, grown with dilution of CH4 with Ar, has the lowest total defect concentration as evidenced from XPS spectral analysis.

Further to these observation by XPS and RS, we attempt to correlate the results obtained by these two techniques. Fig 4 shows the change in defect concentration obtained by XPS analysis as a function of $L_a$ obtained by Raman spectroscopy. As shown in the figure 4, $L_a$ is inversely proportional to the defect concentration which correlates well with the estimates by RS. Also, the FWHM of G band of these samples varies in accordance with assessment made by XPS and provides a qualitative comparison by these two techniques.

As discussed earlier, the defect density obtained from XPS and Raman for samples S1 and S2 agrees with each other. However, in the case of samples S3 and S4, the prediction is bit different in terms of structural quality. For sample S4, XPS studies show better structural quality than S3 which is in straight contradiction to Raman analysis. This difference in prediction by these techniques can be explained as follows: In the first place, XPS is extremely sensitive to surface chemical states and to any deviations in the hexagonal lattice such as a single C vacancy. Such point defects results in altering the BE of C=C bonding in the first to third nearest neighbors of in-plane atoms which can be efficiently detected by XPS. [19] These vacancies introduce chemical shifts and push the BE towards lower values with respect to the C 1s peak of a perfect HOPG. Further, XPS possess high sensitivity only to the surface phenomena and maximum information is obtained only from the topmost few layers with very lower intensity contributed by the subsequent layers. Also, defect bands in XPS spectra do not tend to saturate with increasing defect concentration and allows the estimation of defects to very high levels.

On the other hand, defect band intensity in Raman spectra found to saturate as Raman probes far deep inside the carbon materials, about a micron. Since the scattering cross section in Raman is higher for C=C $sp^2$ bonding than other defective carbon bonding, the spectra dominate with $sp^2$ content and falsely estimate the NGSs as highly oriented. Further, the estimated defect concentration found to give saturated values for highly defective FLG using Raman.[5] This fact is verified experimentally in low energy ion irradiated FLG in which the topmost layers underwent higher damage while the bottom layers survive with their structures intact.[5] Thus, it is evident that the topmost layer of the sample S3, contains planar NGSs grown by diluting $CH_4$ with a gas composition of Ar and $H_2$, undergo severe damage with point defects created during synthesis process. This aspect is reflected correctly as high defect concentration in XPS whereas, Raman and Hall estimates the defect density as low, by the virtue of probing down to the bottom layers wherein the structural quality is found to intact. On the other hand, the samples S1 and S2 are highly defective on surface and interior of the film due to the prevailing rich $H^+$ ions during growth. Thus, Raman and XPS predict defects with similar in the samples S1 and S2.

Generally the NGSs, grown under hydrogen rich plasma as in the cases of S1, S2 and S3, contain large surface defects due to the high etching rate by the atomic hydrogen in the plasma state. On the other hand, the NGSs grown under Ar rich plasma, as in the case of S4, has lesser defect density because $Ar^+$ ions in plasma state enhances the production of $C_2$ radicals which helps to attain higher degree of graphitization.[31] Thus, the defect density estimated by XPS technique on these samples is found to be consistent with the growth mechanism. Hence we conclude based on these studies that although both Raman and XPS techniques can be used for the quantification of defect density in graphitic materials, XPS provide more precise information on point defects and nc-C networks in NGSs.

## Conclusions

In this study, we have investigated defect concentration in few layer NGSs that are grown by plasma enhanced chemical vapor deposition and characterized using XPS and Raman spectroscopy. Herein, a nice one to one correlation in estimating defect density is established between these two techniques, in general. The crystallite size ($L_a$) estimated from Raman spectroscopy is found to have an inverse linear functional relationship with defect density estimated by XPS. Raman spectroscopy found to under estimate the defect density and show saturation in estimating defects in FLG due to presence of structurally intact bottom layers. To conclude, XPS provides an accurate measure of defect concentration and surface chemical state in few layer graphene structures. Hence, XPS is proved to be a superior technique than Raman in estimating the surface defects where technologically crucial applications are based on FLG.

## Acknowledgements

Authors would like to thank Mr.S.R.Polaki and Dr.S.K.Dhara of the department for their fruitful discussion in the course of study. One of the authors, SG, would like to thank Department of Atomic energy, Government of India for research fellowship.